\documentclass[twocolumn,prl,10pt,amsmath,amssymb,nofootinbib,showpacs,superscriptaddress,floatfix]{revtex4-1}

\usepackage{graphicx}
\usepackage{color}
\usepackage[usenames,dvipsnames]{xcolor}
\usepackage[colorlinks=true,linkcolor=Red,citecolor=Green,linktoc=page]{hyperref}
\usepackage{multirow}
\usepackage{float}
\usepackage{flushend}
\usepackage{balance}
\usepackage[varg]{txfonts}
\usepackage{fancyhdr}

\usepackage[usenames,dvipsnames]{xcolor}
\usepackage[colorlinks=true,linkcolor=Red,citecolor=Green,linktoc=page]{hyperref}

\begin{document}

\title{Electrodynamics of ferroelectric films with negative capacitance}
\author{I.\,Luk'yanchuk}
\affiliation{University of Picardie, Laboratory of Condensed Matter Physics,
Amiens, 80039, France} 
\affiliation{Landau Institure for Theoretical
Physics, Moscow, Russia}
\author{A.\,Sen\'{e}}
\affiliation{University of Picardie, Laboratory of Condensed Matter Physics,
Amiens, 80039, France} 
\author{V.\,M.\,Vinokur}
\affiliation{Materials Science Division, Argonne National Laboratory, 9700
S. Cass Ave, Argonne, IL 60439, USA}

\begin{abstract}
We construct a comprehensive theory of the electrodynamic response of
ferroelectric ultra-thin films containing periodic domain textures (PDT)
with $180^{\circ}$ polarization-oriented domains. The focal point of the
theory is the negative-capacitance phenomenon which naturally arises from
the depolarization field induced by PDT. We derive frequency-dependent
dielectric permittivity related to the PDT dynamics across the entire
frequency range. We find the resonance mode of domain oscillations in the
THz spectral band and the singular points in the phase of the reflected THz
beam that are intimately related to the negative capacitance. Our findings
provide a platform for the THz negative capacitance-based optics of
ferroelectric films and for engineering the epsilon-near-zero plasmonic THz
metamaterials.
\end{abstract}

\maketitle

\section*{Introduction}


Functional properties of ferroelectric films and superlattices are
drastically different from those of the bulk materials. Extensive
experimental\,\cite%
{Streiffer2002,Fong2004,Zubko2010,Zubko2012,Hruszkewycz2013,Yadav2016} and
theoretical\,\cite%
{Bratkovsky2000,Bratkovsky2001,Kornev2004,Lukyanchuk2005,DeGuerville2005,Aguado-Fuente2008,Lukyanchuk2009}
studies revealed new physics that emerges from the periodic domain textures
(PDT) stabilized by the strains and the depolarization fields. One of the
most striking features of the PDT is the manifest of the negative
capacitance\,\cite{Zubko2016}, the phenomenon that has been a subject of an
intense recent attention\,\cite%
{Salahuddin2008,Cano2010,Zhirnov2008,Appleby2014,Khan2015,Catalan2015}. The
interest is motivated by both, its fundamental importance, and by its high
potential for technological applications, in particular, as a platform for
novel low-dissipation field-effect transistors.

Another facet of emergent functionalities of ferroelectric films arises due
to the strain-tunability of their multiscale spontaneous polarization
dynamics that provides a wide range of operational frequencies from a few
kilohertz to tens terahertz ($1\,\mathrm{THz}$ $=10^{12}\,\mathrm{Hz}$). The
spectral band $\ 10^{4}\,\mathrm{Hz}-0.3\,\mathrm{THz}$ is covered by the
relaxation dynamics of the domain walls (DWs) and polar clusters. The
far-infrared spectral region of $3-30\,\mathrm{THz}$ is governed by the
soft-mode vibrations of the polar ions. Recent studies suggested that
oscillations of domain walls (DW) in PDT occur in the frequency window of $%
0.3-3\,\mathrm{THz}$\,\cite{Zhang2011,Luk2014,Hlinka2017}, i.e within the
least studied frequency range, referred to as a THz gap.

Our work steps into the breach and demonstrates that the resonant behavior
of oscillations of PDT is the collective effect similar to the standard
plasmonic excitations in metals. We find that this effect is governed by the
stiffness of depolarization field and is a consequence of the negative
capacitance. We calculate the electrodynamic response of PDT and find the
frequency dependence of the effective dielectric permittivity. We
investigate the THz optics of the PDT structure and demonstrate the
existence of the topologically protected phase-singular points of the
absolute darkness in the Fresnel reflection coefficient. These points are a
fingerprint of the negative capacitance phenomenon.

\begin{figure}[tbp]
\center
\includegraphics [width=8cm] {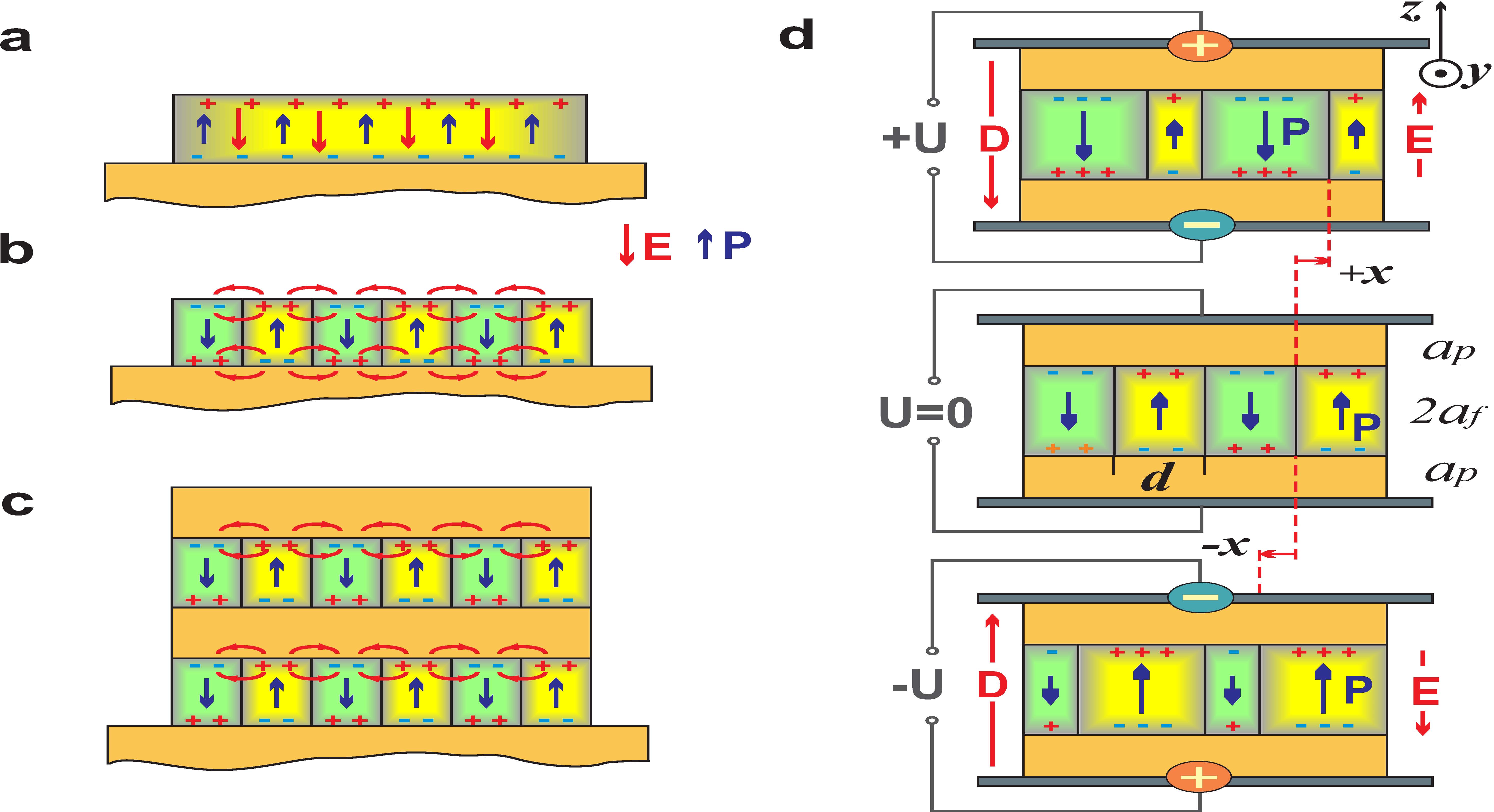}
\caption{ Formation and dynamics of the periodic domain texture (PDT).
(a)\,A single-domain state, the ferroelectric polarization and the
depolarization field are shown by the blue and red arrows, respectively. The
depolarization field traverses across the film. (b)\,The PDT with the
alternating 180$^{\circ}$ polarization domains. The depolarization field is
confined to the surface of the film. (c)\,A sketch of the
ferroelectric-paraelectric superlattice with PDT. (d)\,The change of the PDT
upon varying of the applied field in the ferroelectric-paraelectric system. }
\label{FigOscil}
\end{figure}


Formation of the PDT lowers the energy of the depolarization field induced
by surface depolarization charges that appear at polarization termination
points. Had the polarization maintained the same direction throughout a
slab, see Fig.\,\ref{FigOscil}a, the electrostatic energy stored by the
depolarization field would be proportional to the volume of the whole system
hence huge. Splitting the system into the PDT implies that the surface
depolarization charges form stripes of alternating signs. As a result, the
depolarization field becomes restricted to the near-surface shell, see Fig.\,%
\ref{FigOscil}b, hence drastically diminishing the related electrostatic
energy. Although predicted already in earlier works by Landau\,\cite%
{Landau1935,Landau8} and Kittel\,\cite{Kittel1946} in the context of
ferromagnetic systems, the PDT in ferroelectric films has long been viewed
as unlikely until the recent direct experimental evidences for equilibrium $%
180^{\circ }$ stripe domains in strained ferroelectric thin films of PbTiO$%
_{3}$ (PTO) deposited on the SrTiO$_{3}$ (STO) substrate\,\cite%
{Streiffer2002,Hruszkewycz2013} (as sketched in Fig.\,\ref{FigOscil}b) and
in PTO/STO superlattices\,\cite{Zubko2010,Zubko2012} (Fig.\,\ref{FigOscil}%
c). The observed PDT behaviors appeared to well follow the theoretical
predictions\,\cite%
{Bratkovsky2000,Bratkovsky2001,Kornev2004,Lukyanchuk2005,DeGuerville2005,Aguado-Fuente2008, Lukyanchuk2009}%
. In particular, it was found that the mono-domain $z$-oriented state in the
unscreened films is always unstable with respect the PDT formation.


\begin{figure*}[tbp]
\center
\includegraphics [width=18cm] {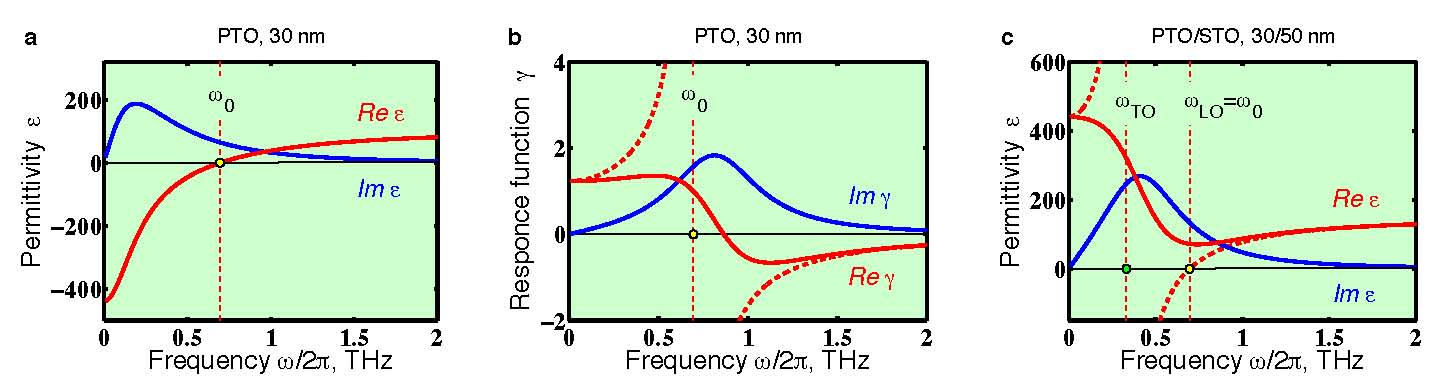}
\caption{ Dynamical properties of the PDT. Real and imaginary parts of the
frequency-dependent response are shown by red and blue colors respectively.
Dashed lines depict the behavior of the system in the absence of
dissipation. (a) \,Dielectric permittivity of the $30\,\mathrm{nm}$ thick
film of the PTO. (b)\,Response function, defined as $\mathbf{P}_{\mathrm{dw}%
\,\protect\omega }=\protect\gamma (\protect\omega )\,\mathbf{D}_{\protect%
\omega }$, for the same film. (c)\,Dielectric permittivity for either the $%
30/50\,\mathrm{nm}$ sandwich of the PTO/STO layers, or for the equivalent
superlattice. }
\label{FigEps}
\end{figure*}

Imagine, however, that surfaces of the slab get short-circuited, for
example, by the metallic electrodes. Then charges would acquire the
capability of arbitrary sliding along the surfaces and flowing between the
electrodes, thus compensating the depolarization field inside the
ferroelectric. As a result, the depolarization energy is nullified, the
domain walls would creep away, and the mono-domain state stabilizes. This
tendency of a ferrolecric with the PDT to self-generating the charge at its
surfaces can be expressed via attributing to the ferroelectric slab a 
\textit{negative} capacitance. One should bear in mind, however, that in
reality, the slab that is not linked to the electrodes, maintains the PDT
since charges cannot propagate across the dielectric and the charge
redistribution described above would not occur.

It is the overscreeing effect due to the depolarization field that lies in
the origin of the negative capacitance. The component of the depolarization
field penetrating the bulk of the ferroelectric slab from the surface is
opposite to the applied field and exceeds it by the absolute value. As a
result, the total field inside the slab is counter-directed to the induced
polarization. Hence the dielectric permittivity is negative. Writing down
the coupled PDT equations of motion and Maxwell equations, we derive the
dynamic frequency-dependent dielectric permittivity of the ferroelectric
films and superlattices containing PDT across the entire frequency range and
find the corresponding resonance mode. We show that its emergence is
intimately related to the low frequency negative dielectric permittivity.

\section*{Results}

\subsection*{Negative capacitance}

We start the description of the domain electrodynamics with the static
limit, and show that PDT in response to the applied field develops the
fascinating negative dielectric permittivity $\varepsilon _{\mathrm{f}}$
and, accordingly, the negative capacitance, $C=\varepsilon _{0}\varepsilon _{%
\mathrm{f}}{S}/(2a_{\mathrm{f}})$, of the ferroelectric layer of surface
area $S$ and thickness $2a_{\mathrm{f}}$. We further demonstrate that the
negative static $\varepsilon _{\mathrm{f}}$ of the PDT is a key element for
striking arising of the resonance mode of DWs oscillations at THz
frequencies.

The applied field is characterized by the induction ${\mathbf{D}}$ that
conserves all across the system, see Fig.\,\ref{FigOscil}d, and generates
the net polarization of the ferroelectric layer, ${\mathbf{P}}\parallel 
\mathbf{z}$ (the electrostatic fields in the ferroelectric slab with domains
are macroscopically averaged). In our ferroelectric film the induction is
related to the electric field through 
\begin{equation}
\mathbf{D}=\varepsilon _{0}\varepsilon _{\Vert }\mathbf{E}+\mathbf{P}_{%
\mathrm{dw}}\,.  \label{ind}
\end{equation}%
The contribution $\mathbf{P_{\mathrm{dw}}}$ stems from the motion of the
domain walls. In the absence of the motion, Eq.\thinspace (\ref{ind})
reduces to standard electrostatic relation between the electric field and
induction. The dielectric constant $\varepsilon _{\Vert }$ is the intrinsic
permittivity of the ferroelectric material along the polarization direction.
The ${\mathbf{P}}_{\mathrm{dw}}$ term reflects the change in the overall
polarization due to the displacement of the DWs altering thus the relative
contribution from the \textquotedblleft up" and \textquotedblleft down"
-oriented domains. This implies the imbalance of the related depolarization
charges of the opposite sign leading to the nonzero average charge density ${%
\sigma }=P_{\mathrm{dw}} $. Thus the depolarization field ${\mathbf{E}}_{%
\mathrm{dep}}=-(\varepsilon _{0}\varepsilon _{\Vert })^{-1}\mathbf{P}_{%
\mathrm{dw}}$ penetrates the bulk of the slab and is directed opposite to
the applied field $\mathbf{D}$. This field, together with the driving field $%
\mathbf{E}_{\mathrm{D}}=(\varepsilon _{0}\varepsilon _{\Vert })^{-1}\mathbf{D%
}$ which would have settled down if DWs were immobile, contributes to the
total field, $\mathbf{E}=\mathbf{E}_{\mathrm{D}}+ \mathbf{E}_{\mathrm{dep}}$%
. Hence the sign of the system's dielectric permittivity, which is defined
as 
\begin{equation}
\varepsilon _{\mathrm{f}}=\frac{D}{\varepsilon _{0}E}\,,  \label{epsf}
\end{equation}
depends on the relative magnitude of the oppositely oriented $\mathbf{E}_{%
\mathrm{D}}$ and $\mathbf{E}_{\mathrm{dep}}$.

Making use of the relation (\ref{ind}), one obtains 
\begin{equation}
\varepsilon _{\mathrm{f}}=\frac{\varepsilon _{\parallel }D}{D-P_{\mathrm{dw}}%
}\,.  \label{epsf1}
\end{equation}
To derive $\varepsilon_f$, we have to know how the polarization related to
the DW displacement depends on the applied field $D$. The calculations
presented in the Appendix result in 
\begin{equation}
\varepsilon _{\mathrm{f}}=\varepsilon _{\parallel }-\frac{{\pi }\varsigma }{%
4\ln 2}\left( \frac{\varepsilon _{\perp }}{\varepsilon _{\parallel }}\right)
^{1/2}\frac{2a_{\mathrm{f}}}{d}\varepsilon _{\parallel }\,,  \label{Epsstat}
\end{equation}
where the first term stems from the positive intrinsic contribution of $%
\mathbf{E}_{\mathrm{D}}$, whereas the second, the negative one, is generated
by the depolarizing field $\mathbf{E}_{\mathrm{dep}}$, reflecting the effect
of the moving DW. The depolarizing term outweights the intrinsic term, hence
the negative dielectric permittivity settles, provided the film thickness, $%
2a_{\mathrm{f}}$ exceeds the domain width, $d$. The equilibrium value of $d$
can be either taken from the experiment\,\cite{Streiffer2002,Zubko2012} or
estimated by the Landau-Kittel square root law\,\cite%
{Landau1935,Kittel1946,Landau8} specifically adapted for ferroelectrics in\,%
\cite{Bratkovsky2000,DeGuerville2005,Catalan2012}, 
\begin{equation}
d\simeq \sqrt{3.53\left( \varepsilon _{\perp }/\varepsilon _{\Vert }\right)
^{1/2}\varsigma\delta\cdot 2a_{\mathrm{f}}}\,.  \label{LK}
\end{equation}%
The resulting typical domain configuration for PTO films is shown in Fig.\,%
\ref{FigOscil}b. The DW thickness, $\delta $, is about 1\thinspace nm~\cite%
{Meyer2002}. In the above formulas $\varsigma =1+\varepsilon _{\mathrm{p}%
}/\left( \varepsilon _{\Vert }\varepsilon _{\perp }\right) ^{1/2}$, and the
values of intrinsic permittivities along and across the polarization
direction, $\varepsilon _{\Vert }$, $\varepsilon _{\perp }$, and of
paraelectric layer, $\varepsilon _{\mathrm{p}}$, are specified below; in the
experimental range $1\leqslant \varsigma \leqslant 4$. If the sandwiching
paraelectric layers possess different permittivities, $\varepsilon _{\mathrm{%
p}}^{+}$ and $\varepsilon _{\mathrm{p}}^{-1}$, the effective parameter $%
\varsigma _{\mathrm{eff}}$ is defined by the relation $\varsigma _{\mathrm{%
eff}}^{-1}=\frac{1}{2}\left[ \varsigma ^{-1}(\varepsilon _{\mathrm{p}%
}^{+})+\varsigma ^{-1}(\varepsilon _{\mathrm{p}}^{-})\right] $. In
particular, if the ferroelectric film is deposited on the metallic substrate
so that $\varepsilon _{\mathrm{p}}^{-}=1$ and $\varepsilon _{\mathrm{p}%
}^{+}=\infty $, one has $\varsigma _{\mathrm{eff}}\simeq 2$. The
polarization profile in this case can be obtained by the image method.

\subsection*{DW oscillations}

To quantify the electromagnetic behavior of PDT, we derive the dynamic
response function $\gamma(\omega)$, defined by the relation $\mathbf{P}_{%
\mathrm{dw}\,\omega }=\gamma (\omega )\,\mathbf{D}_{\omega }$, for the
periodic structure of the domain stripes aligned along the $\mathbf{y}$-axis
which forms in the ferroelectric layer. The polarization axis $\mathbf{z}$
is perpendicular to the film plane, while the in-plane DW motion occurs
along the $\mathbf{x}$-axis, the index $\omega$ indicates the Fourrier
transforms of the respective quantities. We neglect the longitudinal DW
fluctuations which can broad the resonance peak\,\cite{Brierley2014}.

In the harmonic oscillator approximation\,\cite{Sidorkin} the driven
dynamics of DWs is given by%
\begin{equation}
\mu \overset{..}{x}(t)+\eta \overset{.}{x}(t)+k{x}(t)=2P_{\mathrm{s}}E_{%
\mathrm{D}}(t),  \label{Oscil}
\end{equation}%
where $x$ is the coordinate of alternating DW displacements (Fig.\,\ref%
{FigOscil}d), the coefficients $k$, $\mu $ and $\eta $ are calculated per
unit of DW area, and $2P_{\mathrm{s}}E_{\mathrm{D}}(t)$ is the pressure due
to the induction-induced driving field forcing DW to move to flip the
surrounding spontaneous polarization from $-P_{\mathrm{s}}$ to $+P_{\mathrm{s%
}}$.

Equation (\ref{Oscil}) presumes that the DW is rigid and atomically narrow.
However, in the tens-nanometer-thin films the DW broadens significantly due
to depolarization effects in the large range of temperatures (yet the
effective thickness of the DWs remains smaller then the distance between
them as long and the model holds). The broadening is especially pronounced
on approach to the film surface\,\cite{Lukyanchuk2009} and upon the DW
displacement the polarization reverses first at the surface and then
propagates into the interior of the film\,\cite{Zhang2011}. As a result of
the DW broadening, the underlying pinning potential for the DW due to the
periodic atomic structure gets essentially reduced and becomes relevant only
at low temperatures.

The stiffness $k$ is of a purely electrostatic origin and, therefore, is not
too sensitive to the explicit shape of DW. We calculate it as a coefficient
at the restoring force that pushes the domain walls back in order to reduce
the depolarization field, $E_{\mathrm{dep}}=-\Delta P_{\mathrm{s}%
}/\varepsilon _{0}=-2(x/\varepsilon _{0}d)P_{\mathrm{s}}$, arising upon the
displacment of DW from the equilibrium. This field is induced by the extra
depolarization surface charges, $\sigma =\pm \Delta P_{\mathrm{s}}$,
appearing due to the spontaneous polarization excess, $\Delta P_{\mathrm{s}%
}=2(x/d)P_{\mathrm{s}}$, and is directed antiparallel to $\Delta P_{\mathrm{s%
}}$. We evaluate the depolarization energy of the system as the
electrostatic energy of the dielectric slab with permittivity $\varepsilon
_{\Vert }$ and with surface charges $\sigma =2({x}/{d})P_{\mathrm{s}}$ as $%
W=\left( 2a_{\mathrm{f}}\sigma ^{2}/2\varepsilon _{0}\varepsilon _{\Vert
}\right) S$, where $S$ is the surface area. The corresponding energy per
unit area of the displaced DW is $w=(d/2Sa_{\mathrm{f}})W$. Relating it with
the harmonic oscillator stiffness energy $kx^{2}/2$, we find 
\begin{equation}
k=\frac{4P_{\mathrm{s}}^{2}}{\varepsilon _{0}\varepsilon _{\Vert }d}g,\quad
g\simeq 1-\frac{4\ln 2}{\pi \varsigma }\left( \frac{\varepsilon _{\parallel }%
}{\varepsilon _{\perp }}\right) ^{1/2}\frac{d}{2a_{\mathrm{f}}}.
\label{stiffness}
\end{equation}%
The factor $g\leq 1$, calculated in the Appendix, makes the above
qualitative consideration more precise, taking into account the non-uniform
part of the depolarization field near the surface caused by the stepwise
distribution of the depolarization surface charges. As follows from (\ref%
{stiffness}) in very thick samples, these corrections vanish and $%
g\rightarrow 1$. In realistic cases that we considered here, namely, the PTO
films with $2a_{\mathrm{f}}\simeq 10-30$\,nm, it varies from 0.4 to 0.9.


The effective DW mass, $\mu $, and the viscosity, $\eta $, in Eq.~(\ref%
{Oscil}) are related to the motion of the material-constituent polar ions
during the displacement of the DWs\,\cite{Kittel1951} and with the
piezoelectric effect of the time-varying depolarization field\,\cite%
{Sidorkin}. The magnitude of $\mu $ was calculated in \textit{ab-initio}
simulations of DW dynamics in the sub-THz range\,\cite{Zhang2011}. We
propose an interpolation formula $\mu \,\mathrm{[kg/m}^{2}\mathrm{]}\simeq
1.3\sqrt{2a_{\mathrm{f}}\,\mathrm{[nm]}}\times 10^{-9}$, which fits well the
numerical result. Viscosity, $\eta$, is expressed via damping factor of DW
motion, $\Gamma =\eta /\mu $. Since the consistent theory for $\Gamma $ is
not available, we chose the one that appears in the measurements of the
soft-mode relaxation of polar ions.

\subsection*{THZ resonance}

In what follows we calculate the complex dynamic permittivity of PDT, $%
\varepsilon _{\mathrm{f}}(\omega )$. The resulting plots of the real, $\Re
\/\varepsilon _{\mathrm{f}}(\omega )$, and imaginary, ${\Im }$\/$\varepsilon
_{\mathrm{f}}(\omega )$, parts of the dispersion $\varepsilon _{\mathrm{f}%
}(\omega )$ for 30\,nm thick film of PTO are shown in Fig.\,\ref{FigEps}a.

The reaction of the ferroelectric layer to the time-dependent applied field $%
\mathbf{D}(t)=\int d\omega \,\mathbf{D}_{\omega }e^{-i\omega t}$ is
characterized by the linear response function, $\gamma (\omega )$. Solving
Eq.\,(\ref{Oscil}) by Fourier transformation, we find: 
\begin{equation}
\gamma (\omega )=\frac{g^{-1}\,\omega _{0}^{2}}{\omega _{0}^{2}-\omega
^{2}-i\Gamma \omega },\,\,\,\,\omega _{0}=\left( \frac{4P_{\mathrm{s}}^{2}g}{%
\mu d\varepsilon _{0}\varepsilon _{\Vert }}\right) ^{1/2}\, ,  \label{gomega}
\end{equation}%
where the characteristic oscillation frequency is the usual $\omega
_{0}=\left( k/\mu \right) ^{1/2}$ with $k$ and $\mu$ related to the system
parameters as discussed above. When deriving Eq.\,(\ref{gomega}), we took
into account that $\mathbf{P}_{\mathrm{dw}\,\omega }=2P_{\mathrm{s}}\,\left(
x_{\omega }/d\right) \mathbf{z}$ and $\mathbf{D}_{\omega }=\varepsilon
_{0}\varepsilon _{\Vert }\mathbf{E}_{\mathrm{D}\,\omega }$.

\begin{figure}[b!]
\par
\begin{center}
\includegraphics [width=8cm] {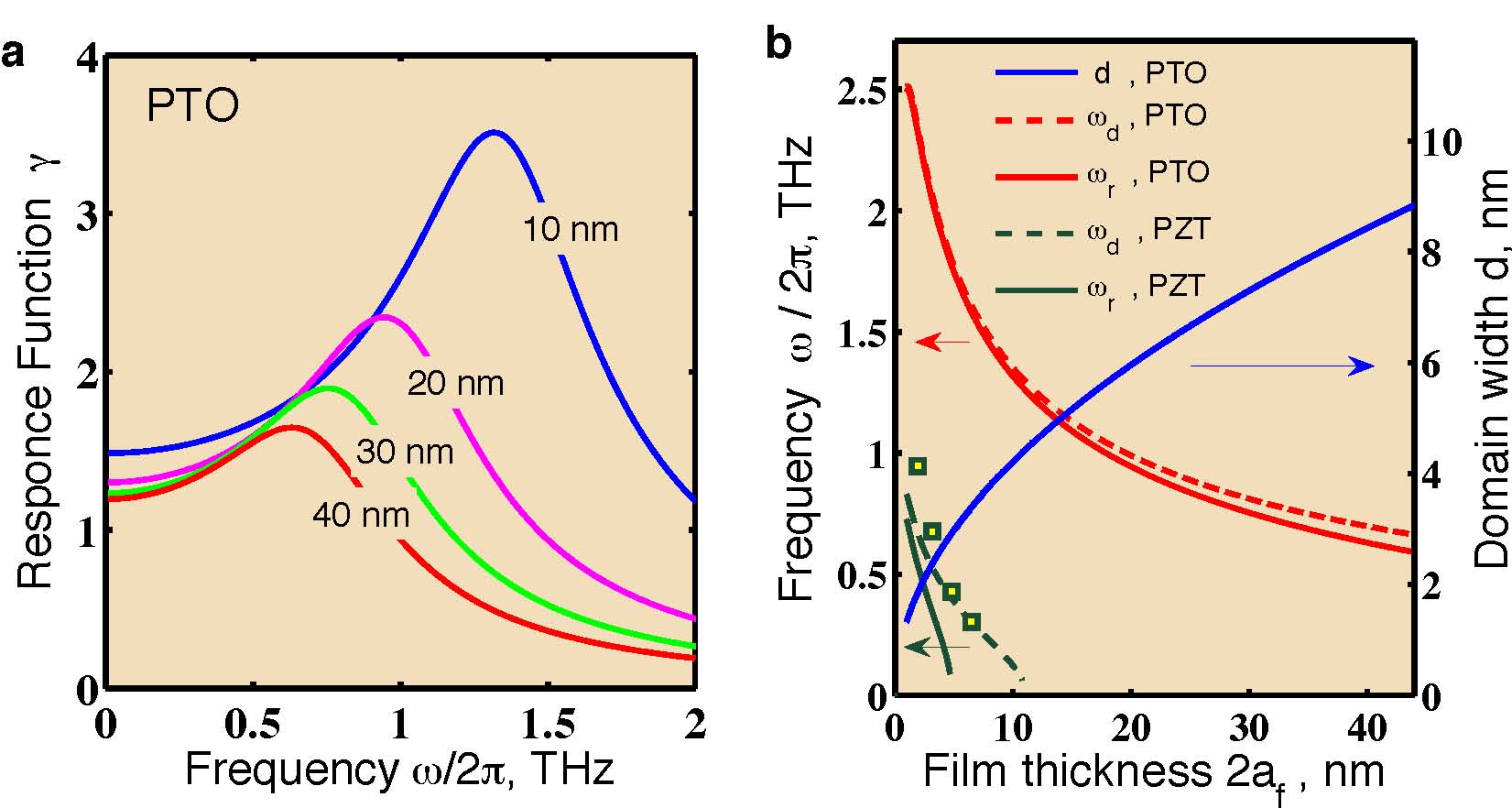}
\end{center}
\caption{Resonance in the PDT. (a)\,Resonance behavior of the amplitude of
the response function $|\protect\gamma (\protect\omega )|$ for different
film thickness of the PTO. (b) \,Dependencies of the domain resonance
frequency, $\protect\omega_\mathrm{r}$, of the damping frequency, $\protect%
\omega_\mathrm{d}$, and of the domain width, $d$, upon the film thickness.
The solid red line stands for $\protect\omega_\mathrm{r}$ of the PTO, the
dashed red line shows $\protect\omega_\mathrm{d}$ of the PTO. The solid and
dashed green lines show $\protect\omega_\mathrm{r}$ and $\protect\omega_%
\mathrm{d}$, respectively, for PZT. Green square symbols display the results
of the \textit{ab-initio} simulations for the PZT\,\protect\cite{Zhang2011}.
The blue solid line displays the behavior of the domain width, $d$.}
\label{FigRes}
\end{figure}

Shown in Fig.\,\ref{FigEps}b are the real and imaginary parts of $\gamma
(\omega )$ for our sample exhibiting a characteristic Lorentzian-type form.
The latter is a legacy of a singularity at $\omega _{0} $, which would have
been exhibited by a system without the dissipation, illustrated by the
dashed lines. The emergence of the resonance response to the applied field
is clearly demonstrated by the frequency dependencies of the absolute value
of the response function, $|\gamma (\omega )|$, for different film
thicknesses, see Fig.\,\ref{FigRes}a, and reaching its maximum at the
resonance frequency $\omega _{\mathrm{r}}^{2}=\omega _{0}^{2}-\Gamma ^{2}/2$%
. The dependence of $\omega_\mathrm{r}$ upon the film thickness, $2a_\mathrm{%
f}$, is shown in Fig.\,\ref{FigRes}b. Shown by the blue line in the same
panel is the Landau-Kittel dependence of the domain width, $d$, on the film
thickness given by Eq.\,(\ref{LK}) .

Equation\,(\ref{gomega}) enables optimization of the materials parameters
and the film thickness to make sure that $\omega _{\mathrm{r}}$ falls within
the desired THz frequency range. In particular, for the strained films of
PTO with the high spontaneous polarization, $P_{s}\simeq 0.65\,\mathrm{C\,m}%
^{-2}$, see\,\cite{Pertsev2003}, relatively low permittivities $\varepsilon
_{\Vert }\simeq 100$, $\varepsilon _{\perp }\simeq 30$, soft mode damping
factor $\Gamma \simeq 20\,\mathrm{cm}^{-1}$ ($0.6\,\mathrm{THz}$)\thinspace 
\cite{Hlinka2011} and $\mu $ defined as above, the resonance frequency $%
\omega _{r}$ decreases and spans the range from $1.5$ to $0.75\,\mathrm{THz}$
when $2a_{\mathrm{f}}$ increases from $10$ to $40\,\mathrm{nm}$. 
%
We find the damping frequency, $\omega _{\mathrm{d}}^{2}=\omega
_{0}^{2}-\Gamma ^{2}/4$, of the attenuated oscillations of domains in PTO, $%
x(t)=x_{0}e^{-(\Gamma /2)\,t}\sin \omega _{\mathrm{d}}t$, which is slightly
larger then $\omega _{\mathrm{r}}$, see Fig.\,\ref{FigRes}b. Remarkably, our
formulas perfectly describe the results of \textit{ab-initio} simulations of
DWs oscillations in PZT free-standing ultrathin films\,\cite{Zhang2011}. The
calculated damping frequency, $\omega _{\mathrm{d}}$, is shown by the dashed
line in the Fig.\,\ref{FigRes}b, the symbols display the results of
simulations. Here we used the following parameters for PZT: $P_{\mathrm{s}%
}\simeq 0.40\,\mathrm{C\,m}^{-2}$\,\cite{Pertsev2003}, $\varepsilon _{\Vert
},\varepsilon _{\perp }\simeq 350$, $\Gamma \simeq 27\mathrm{cm}^{-1}$ ($%
0.8\,\mathrm{THz}$)\thinspace \cite{Buixaderas2010} and the same value for
DW mass, $\mu$. At the same time, the resonance frequency of PZT, $\omega _{%
\mathrm{r}}$, drops rapidly with the thickness and vanishes above
4\thinspace nm. The data in this paragraph are given for ceramic samples.
The values for strained films can be slightly different.

Making use of the relation $\gamma (\omega )=1-\varepsilon _{\Vert
}/\varepsilon _{\mathrm{f}}(\omega )$ that follows directly from Eq.\,(\ref%
{ind}), we obtain the dynamic permittivity of a ferroelectric layer 
\begin{equation}
\varepsilon _{\mathrm{f}}(\omega )=\frac{\varepsilon _{\Vert }}{1-\gamma
(\omega )}=\varepsilon _{\Vert }\frac{\omega _{0}^{2}-\omega ^{2}-i\Gamma
\omega }{\left( 1-g^{-1}\right) \,\omega _{0}^{2}-\omega ^{2}-i\Gamma \omega 
},  \label{EpsFw}
\end{equation}%
with $1-g^{-1}<0$. At small $\omega $ this yields the negative static
permittivity described by Eq.\,(\ref{Epsstat}). We plot the real, $\Re
\/\varepsilon _{\mathrm{f}}(\omega )$, and imaginary, ${\Im }$\/$\varepsilon
_{\mathrm{f}}(\omega )$, parts of complex dispersion, $\varepsilon _{\mathrm{%
f}}(\omega )$, calculated for 30\thinspace nm thick film of PTO in
paraelectric environement, see Fig.\,\ref{FigEps}a. At low frequencies $\Re
\varepsilon _{\mathrm{f}}(\omega )$ remains negative. At high frequencies
the oscilations of DWs freeze out, and therefore $\varepsilon _{\mathrm{f}%
}(\omega )$ should become positive and equal to intrinsic permettivity, $\Re
\/\varepsilon _{\mathrm{f}}(\infty )=\varepsilon _{\Vert }>0$. At frequency $%
\omega =\omega _{0}$, the real part of permittivity turns zero, $\Re
\/\varepsilon _{\mathrm{f}}(\omega _{0})=0$. This behavior of PDT
permittivity resembles that of a metal, having the negative real part of
permittivity below the \textit{plasma frequency} at which it becomes equal
to zero. Accordingly, this will lead to the plasma resonance at $\omega
=\omega _{0}$ in a PDT as a response to the driving field $\mathbf{D}$.

The discovered resonance mode gives rise to the Drude-Lorentz frequency
response of a biased capacitor consisting of the ferroelectric layer (of
thickness $2a_{\mathrm{f}}$) sandwiched between paraelectric buffer layers
(of thickness $a_{\mathrm{p}}$ and permittivity $\varepsilon _{\mathrm{p}}$
each). The same response appears in $2a_{\mathrm{f}}/2a_{\mathrm{p}}$
ferroelectric-paraelectric superlattice. The effective permittivity,
calculated for the system of in-series capacitors, $\varepsilon _{\mathrm{tot%
}}^{-1}(\omega )=\alpha _{\mathrm{p}}\varepsilon _{\mathrm{p}}^{-1}+\alpha _{%
\mathrm{f}}\varepsilon _{\mathrm{f}}^{-1}(\omega )$, results from the
interplay of the positive, paralectric, and negative, ferroelectric,
contributions and remains positive at small $\omega $. Here $\alpha _{%
\mathrm{p}}=a_{\mathrm{p}}/(a_{\mathrm{p}}+a_{\mathrm{f}})$ and $\alpha _{%
\mathrm{f}}=a_{\mathrm{f}}/(a_{\mathrm{p}}+a_{\mathrm{f}})$ are the relative
weights of the respective layers. That it is the paraelectric contribution
that dominates the behavior of the effective permittivity at $\omega =0$, is
inherently related to the very emergence of the domains. Namely, it is
related to the fact that the thickness of the paraelectric buffer, $a_{%
\mathrm{p}}$, is larger then the domain width, $d$, and thus domain
depolarization stray fields do not interact with electrodes\,\cite{Mokry2004}%
. The frequency dependence of the $\varepsilon _{\mathrm{tot}}(\omega )$ for
PTO/STO $30/50$ $\mathrm{nm}$ system is shown in Fig.\,\ref{FigEps}c. One
sees (the detail of calculations are given further) that $\Re \,\varepsilon
_{\mathrm{tot}}(\omega )$ maintains a positive value for low frequencies and
reflects the presence of the oscillatory mode at $\omega $ slightly less
then $\omega _{0} $. As before, the dashed line depicts the behavior of $\Re
\,\varepsilon _{\mathrm{tot}}(\omega )$ in the absence of dissipation. We
thus introduce a new metamaterial operating in the THz frequency-range $%
\varepsilon _{\mathrm{tot}}(\omega )$ tunable by the variation of ratio $2a_{%
\mathrm{f}}/2a_{\mathrm{p}}$.

\begin{figure*}[tbp]
\par
\begin{center}
\includegraphics [width=14cm] {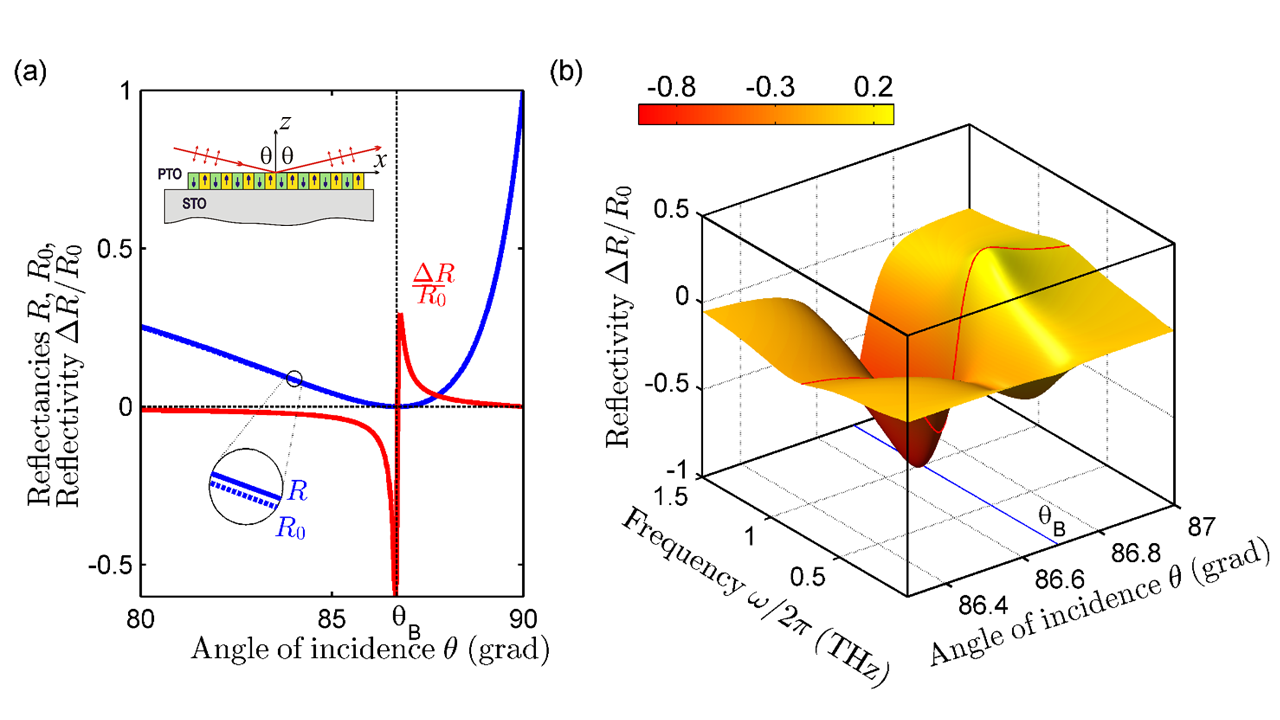}
\end{center}
\caption{Reflection characteristics of the PDT. The $p$-polarized THz beam
reflects from the $2a_\mathrm{f}\simeq30\,\mathrm{nm}$-thick PTO film
deposited on the STO dielectric substrate. (a) The dependence of the
reflectance of the film on the substrate, $R$, and the dependence of the net
reflectance of the substrate, $R_0$, on the incidence angle, $\protect\theta$%
, calculated at the PDT oscillation frequency $\protect\omega_\mathrm{0}/2%
\protect\pi \simeq 0.75\,\mathrm{THz}$ are shown by the blue solid and
dashed lines correspondingly. The difference between the both is tiny, $%
|\Delta R|=|R_0-R| \lesssim 10^{-3}$, and in the figure scale is seen only
under magnification. At variance, the reflectivity, $\Delta R/R_0$,
experiences the drastic variation of about unity and clearly displays the
Lorentz-like spike at the Brewster angle $\protect\theta _{\mathrm{
\scriptscriptstyle B}}\simeq 86.7^{\circ}$ where $R_0$ almost vanishes. This
spike is a fingerprint of the PDT resonance and can be used for its
experimental identification. The inset shows the geometry of the experiment.
(b)\,The 3D color plot of the reflectivity, $\Delta R /R_0$, as function of
the frequency and the incident angle in the immediate vicinity of the
Brewster angle, $\protect\theta_\mathrm{B}$, marked by thin blue line. The
red line corresponds to the Lorentz-spike-like reflectivity behaviour at the
PDT oscillation frequency, $\protect\omega_0$. This spike smoothens upon the
deviation from $\protect\omega_0$.}
\label{FigReflection}
\end{figure*}

\begin{figure*}[tbp]
\par
\begin{center}
\includegraphics [width=14cm] {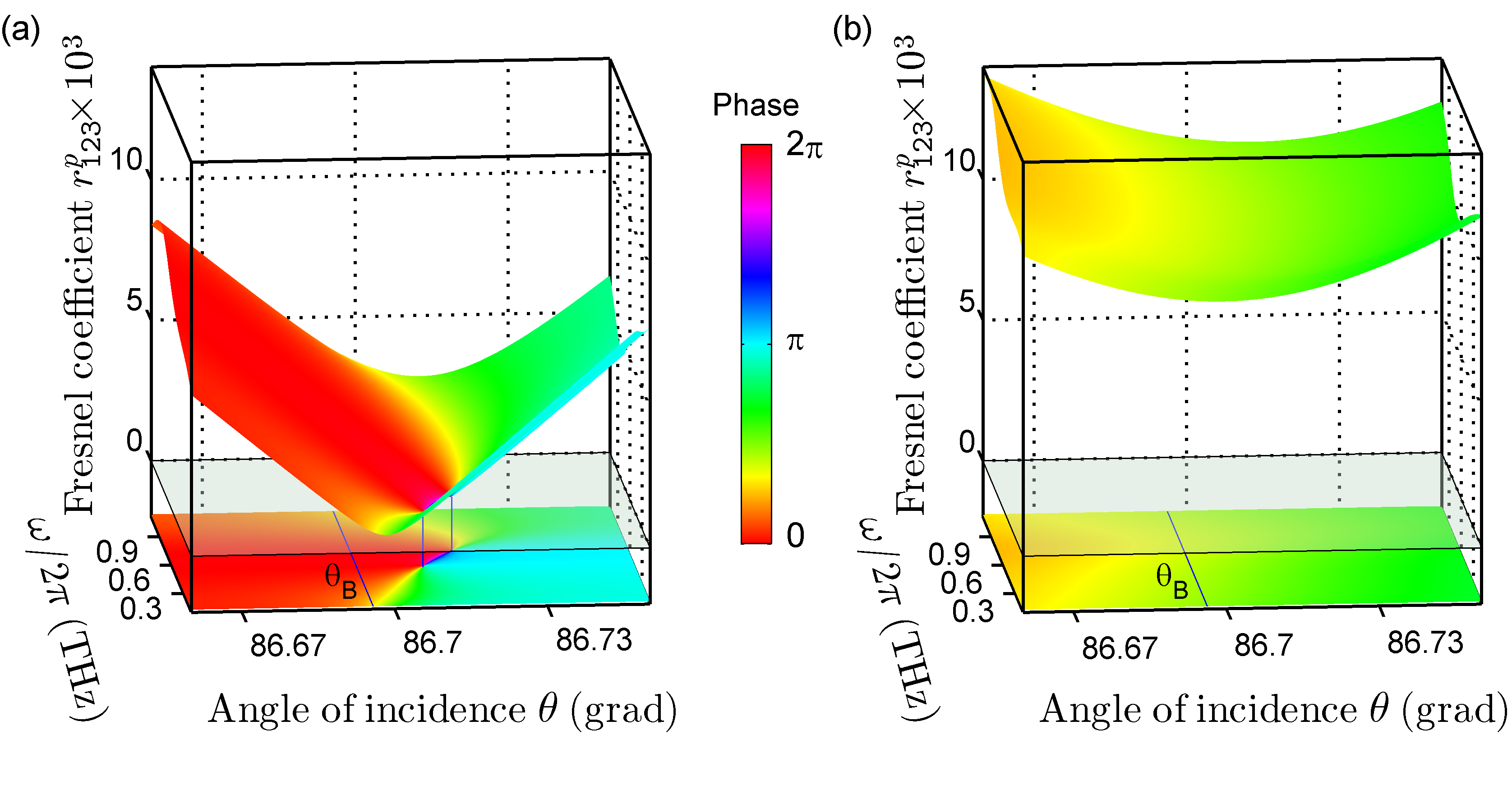}
\end{center}
\caption{The complex Fresnel reflection coefficient, $r_{123}^p$, of the
PDT. (a)\,The 3D color plot of the phase and the amplitude of $r_{123}^p$ as
function of the frequency, $\protect\omega$, and the angle of incidence, $%
\protect\theta$, for the $2a_\mathrm{f}\simeq30\,\mathrm{nm}$-thick PTO film
deposited on the transparent substrate. The substrate permittivity $\protect%
\varepsilon_\mathrm{p}=300$ is real with the vanishingly small imaginary
part. Two topologically protected phase singularity dark points, at which $%
|r_{123}^p|=0$, appear in the vicinity of the Brewster angle $\protect\theta_%
\mathrm{B}$ at frequencies where $\protect\varepsilon_\mathrm{f}=0$ and
where $\protect\varepsilon_\mathrm{f}\simeq -\protect\sqrt{\protect%
\varepsilon_\mathrm{p} \protect\varepsilon_\perp}$. These singular points
and the inversion of the phase rotation between them that came out in blue
is the fingerprint of the negative permittivity in the low-frequency region
at $\protect\omega<\protect\omega_0$. (b) The same as (a) but for the finite
transparency STO substrate with $\protect\varepsilon_\mathrm{p}=300+10i$.
The singular phase dark points merge and annihilate. The color legend refers
to both panels (a) and (b). The color maps of the phase distribution are
shifted down from $|r_{123}^{p}|=0$ plain for convenience. }
\label{FigFresnel}
\end{figure*}


\subsection*{Optics}

The effective permittivity of a ferroelectric/paraelectric layered system is
calculated as $\varepsilon _{\mathrm{tot}}^{-1}(\omega )=\alpha _{\mathrm{p}%
}\varepsilon _{\mathrm{p}}^{-1}+\alpha _{\mathrm{f}}\varepsilon _{\mathrm{f}%
}^{-1}(\omega )$ and can be conveniently written in a standard form: 
\begin{equation}
\varepsilon _{\mathrm{tot}}(\omega )=\frac{\omega _{\mathrm{\mathrm{%
\scriptscriptstyle LO}}}^{2}-\omega ^{2}-i\Gamma \omega }{\omega _{\mathrm{%
\mathrm{\scriptscriptstyle TO}}}^{2}-\omega ^{2}-i\Gamma \omega }\varepsilon
_{\mathrm{tot}}(\infty )\,,  \label{EpsTotw}
\end{equation}%
where the characteristic frequencies $\omega_{\mathrm{\scriptscriptstyle LO}}
$, $\omega_{\mathrm{\scriptscriptstyle TO}}$ are the analogues of the
longitudinal and traversal oscillation frequencies in the polar mode
spectroscopy. The former one, is the DW oscillation frequency, $\omega_{%
\mathrm{\scriptscriptstyle LO}}=\omega_0$. The latter frequency can be
obtained from the relation $\omega _{\mathrm{\mathrm{\scriptscriptstyle TO}}%
}^{2}/\omega _{\mathrm{\mathrm{\scriptscriptstyle LO}}}^{2}=\varepsilon _{%
\mathrm{tot}}(\infty )/\varepsilon_{\mathrm{tot}}(0)$ that is the analog of
the Lyddane-Sachs-Teller relation. Here $\varepsilon_{\mathrm{tot}%
}^{-1}(0)=\alpha_{\mathrm{p}}\varepsilon_{\mathrm{p}}^{-1}+\alpha_{\mathrm{f}%
}\varepsilon_{\mathrm{f}}^{-1}$ and $\varepsilon_{\mathrm{tot}}^{-1}(\infty
)=\alpha_{\mathrm{p}}\varepsilon_{\mathrm{p}}^{-1}+\alpha_{\mathrm{f}%
}\varepsilon_{\Vert }^{-1}$. Equation\thinspace(\ref{EpsTotw}) describes the
dispersion curve in Fig.\,\ref{FigEps}c. It clearly shows that the obtained
PDT resonance mode freezes out when the frequency exceeds $\omega_0$ and
what is left are the molecular vibrations including the polar soft mode
oscillations.

The derived dispersion relation for the dielectric permittivity provides a
foundation for the description of the THz optical properties of a
ferroelectric film with PDT, in particular, for finding its complex
refraction index, $n(\omega)=\sqrt{\varepsilon(\omega)}$. This enables to
devise an approach for inferring the resonance properties and negative
permittivity effects from the optical measurements in the sub(THz) frequency
band. Importantly, in the systems that we address, the typical thicknesses
of the ferroelectric layer, $2a_{\mathrm{f}}\simeq 10\div 40\,\mathrm{nm}$
and the PDT period, $2d$, are both less then the wavelength of THz
radiation, $\lambda \simeq 1\div 0.15\,\mathrm{mm}$ for $0.3\div 2\,\mathrm{%
THz}$. Hence the THZ wave will interact with the system hosting the PDT as
with an 
ultra-thin dielectric film endowed with the anisotropic dielectric
permittivity, the latter having the in- and out of plane components $%
\varepsilon_\perp$ and $\varepsilon _\mathrm{f}(\omega)$ respectively.

\subsection*{Reflection}

We consider the situation where the incident THz beam gets reflected from
the dielectric substrate, STO, with dielectric permittivity $\varepsilon_%
\mathrm{p}=300+10i$ in the sub-THz frequency range, passing twice through
the PTO ferroelectric film with PDT. To ensure the interaction of the beam
with the PDT polarization, the electric field of the light should be
confined to the plane of incidence, hence the p-wave polarization geometry
should be used. The experimental setup is shown in the inset in Fig.\,\ref%
{FigReflection}a. The calculated reflectance, $R$, of such a system is shown
in the main panel of Fig.\,\ref{FigReflection}a as function of the angle of
incidence, $\theta$, by the solid blue line. The plot illustrates a typical
p-wave reflectance with the amplitude that nearly vanishes in the vicinity
of the Brewster angle of the substrate, $\theta _{\mathrm{\scriptscriptstyle{%
B}}}\approx \arctan \left( \Re \mathrm{\,}\varepsilon _{\mathrm{p}%
}\right)^{1/2}\simeq 86.7^{\circ }$. For comparison, shown in Fig.\,\ref%
{FigReflection}a by the blue dashed line is also the net contribution of the
substrate. Note that because of the very small thickness of the film, it is
practically indistinguishable from the reflectance of the whole system. The
difference between them, $\Delta R=R_0-R \simeq 10^{-3}$, is seen only under
a magnification. Thus, to ensure the efficient filtering of the resonance
PDT signal from the background substrate contribution, one has to focus
instead on the relative variance of the signals difference, $\Delta R/R_0$,
usually referred to as to reflectively and used to detect the tiny molecular
mono-layers adsorbed on the substrate\,\cite{Tolstoy2003}, see Fig.\,\ref%
{FigReflection}.

An advantage of using the reflectivity is that the value of the substrate
reflectance, $R_0$, almost vanishes at $\theta _{\mathrm{\scriptscriptstyle{B%
}}}$ and the use of the ratio $\Delta R/R_0$ unravels the net contribution
of the ultra-thin film \textit{per se}. Shown by the solid red line, the
reflectivity of our system does experience the drastic variation of about
unity and clearly displays the Lorentz-like spike in vicinity of $\theta _{%
\mathrm{\scriptscriptstyle B}}$. This spike is a fingerprint of the PDT
resonance and can be used for its experimental identification. The complete
qualitative behavior of the resonance feature is illuminated by the
three-dimensional color plot of $\Delta R /R_0$, as function of the
frequency and the angle of incidence (Fig.\,\ref{FigReflection}b). The
Lorentz-spike sharp behaviour gets smoothed upon the deviation from the
resonance frequency.

The reflectancies $R$ and $R_0$ are calculated as a square of the magnitude
of the corresponding Fresnel reflection coefficients, $R=|r^p_{123}|^2$ and $%
R_0=|r^p_{13}|^2$. Here index $p$ refers to the beam polarization and
indices 1, 2, and 3 correspond to the ordering of the media: the air, the
film, and the substrate, respectively.

The Fresnel reflection coefficient of the substrate-deposited film is given
by the expression\,\cite{Tolstoy2003} 
\begin{equation}
r_{123}^{p}=\frac{r_{12}^{p}+r_{23}^{p}e^{2i\beta }}{%
1+r_{12}^{p}r_{23}^{p}e^{2i\beta }},  \label{Reflectance}
\end{equation}%
where 
\begin{equation}
r_{ij}^{p}=\frac{\varepsilon _{j}\xi _{i}-\varepsilon _{i}\xi _{j}}{%
\varepsilon _{j}\xi _{i}+\varepsilon _{i}\xi _{j}}, \qquad i,j=1,2,3
\label{Fresnel}
\end{equation}%
are the generalized Fresnel reflection coefficients between two adjacent
media and $\beta =2\pi \left( 2a_{f}/\lambda \right) \xi _{2}$ is the phase
shift of the electromagnetic wave after a single pass through the film. The
generalized indices of refraction, $\xi _{i}$, are defined via the angle of
incidence, $\theta_1$, and dielectric permittivities of the media, $%
\varepsilon_i$, as 
\begin{gather}
\xi _{1}=\sqrt{\varepsilon _{1}}\cos \theta _{1},\quad \xi _{2}=\left(
\varepsilon _{\perp }-\frac{\varepsilon _{\perp }}{\varepsilon _{2}}%
\varepsilon _{1}\sin ^{2}\theta _{1}\right) ^{1/2} \\
\xi _{3}=\left( \varepsilon _{3}-\varepsilon _{1}\sin ^{2}\theta _{1}\right)
^{1/2}\,.  \label{Refraction}
\end{gather}
While the refraction indices $\xi_1$ and $\xi_3$ represent the customary
properties of the air, $\varepsilon_1=1$, and substrate, $%
\varepsilon_3=\varepsilon_\mathrm{p}$ endowed with the isotropic
permittivities, the targeted electrodynamic response of PDT with negative
capacitance is encoded in the anisotropic refraction index $\xi_2$ that is
deduced from\,\cite{Schopper1952}. The latter grasps both the transversal
permittivity of the film, $\varepsilon_{\perp}$, and the effective
longitudinal permittivity of PDS, $\varepsilon _{2}=\varepsilon _{\mathrm{f}%
}(\omega)$.

The described approach identifies the PDT resonance behavior via detecting
the characteristic Lorentz-like spike features in the angular and frequency
dependencies of the reflectance.  However, the developed theory of the
electromagnetic response equips us with the extraordinary more advanced
technique. Namely, the resonance point $\omega_0$ where $\varepsilon_\mathrm{%
f}(\omega)$ changes the sign, is found, with the great precision from the
phase map of the Fresnel coefficient $r_{123}^{p}$. The latter (unlike the
real-value reflectance $R=|r_{123}^{p}|^2$, where the phase information is
lost) is a complex quantity which is measured, for example, by the
phase-resolved ellipsometry technique\,\cite{Marsik2016}.

To develop a feasible protocol for observation of the frequency domain
endowed with negative $\varepsilon_\mathrm{f}$ let us describe the
dependence $r_{123}^{p}(\theta)$ near the Brewster angle in the system with
the ideally transparent substrate with $\Im \varepsilon_\mathrm{p}=0$. If
the film were absent, the Fresnel coefficient would have been real and
changed the sign upon passing zero at $\theta=\theta_{\mathrm{%
\scriptscriptstyle B}}$. Placing the film on the substrate, gives the
contribution, albeit very small, that pulls $r_{123}^{p}$ into a complex
plane, where the trajectory $r_{123}^{p}(\theta)$ circumflexes the zero from
the positive- to negative semi-axis acquiring the phase $\pm \pi$. Equations
(\ref{Reflectance})-(\ref{Refraction}) offering a full description of the
electromagnetic response show that the direction (clock- vs. anticlock-wise)
of the phase rotation is determined by the sign of $\varepsilon_\mathrm{f}$.
The frequency $\omega_0$ at which $\varepsilon_\mathrm{f}$ changes the sign
corresponds, therefore, to a particular point in the $(\theta,\omega)$-plane
where phase acquires a factor of $2\pi$ when making a close loop around it,
and the Fresnel coefficient, $r_{123}^{p}$, vanishes. These zeros of the
complex function, $r_{123}^{p}(\theta,\omega)$, where the intensity of the
reflected light is zero are often called `points of absolute darkness'\,\cite%
{Kravets2013}. They are stable with respect to small variations of
parameters hence topologically protected, and we refer to them as to
topological darkness points.

Figure\,\ref{FigFresnel}a displays the 3D color plot of the phase and
amplitude of the Fresnel reflection coefficient, $r_{123}^{p}$, of the PDT
in the 30nm PTO film as a function of an incident angle and frequency for
the case of transparent substrate with $\varepsilon_\mathrm{p}=300$, derived
from Eqs.\,(\ref{Reflectance})-(\ref{Refraction}). Two topological dark
points resulting from the negative permittivity are clearly visible at the
manifold depicting $r_{123}^{p}(\theta,\omega)$. The higher-frequency point
A corresponds to the discussed above sign change of $\varepsilon_\mathrm{f}$%
. The second, lower-frequency one, appears when $\varepsilon_\mathrm{f}%
\simeq -\sqrt{\varepsilon_\mathrm{p} \varepsilon_\perp}$. This manifestly
illustrates the topological protection effect: the topological darkness
points are not smoothed but merely shifted by dissipation. Figure\,\ref%
{FigFresnel}b shows the the same phase plot for the PTO on a real substrate
STO with $\varepsilon_\mathrm{p}=300+10i$. One sees that if the dissipation
is sufficiently large, the darkness points merge and cease to exist, i.e.
the sharp Brewster-angle singularities in the phase plot get smoothened.

The obtained results and recent measurements of the negative capacitance in
PDT\,\cite{Zubko2016} open a new area of research in the field of the THz
optics in ferroelectric materials. Exciting opportunities open in the area
of the plasmonic epsilon-near-zero (ENZ) metamaterials having the unique
property that the electromagnetic wave propagates with almost no phase
advance. Although such materials have been made artificially in the
microwave, visible and far-infrared spectral ranges\,\cite{Maas2003}, the
engineered ENZ structures in the THz spectral range have not been explored
so far.

\subsection*{Acknowledgments}

The work was supported by the U.S. Department of Energy, Office of Science,
Materials Sciences and Engineering Division (VV) and European
FP7-MC-ITN-NOTEDEV and H2020-MC-RISE-ENGIMA actions (IL).

\section*{APPENDIX}

\begin{figure}[bth]
\center
\includegraphics [width=6cm] {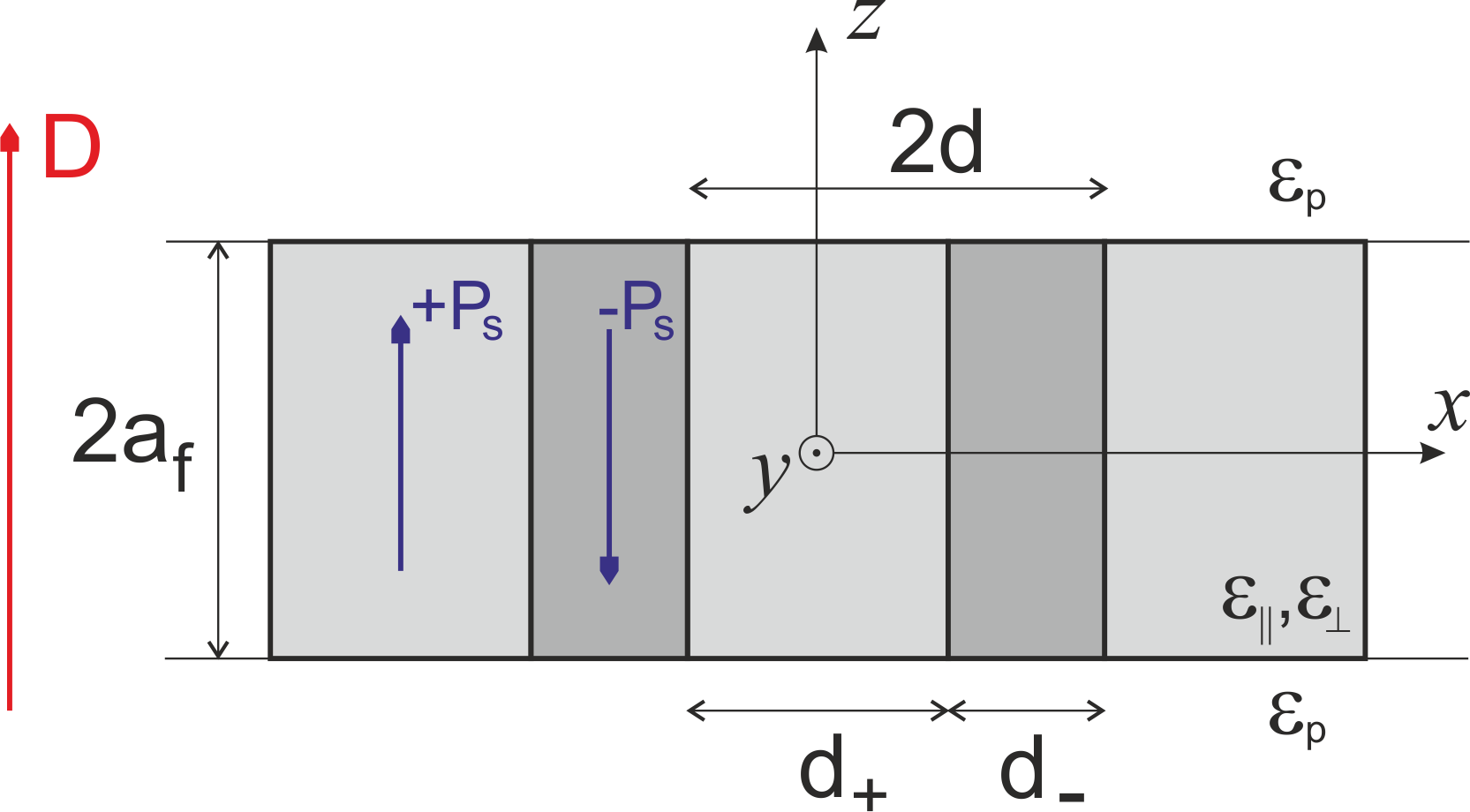}
\caption{ The geometry of the system. }
\label{FigSuplGeom}
\end{figure}

Calculations of the electrostatic properties of PDT in the applied field $%
\mathbf{D}$ generalize the zero-field Landau-Kittel calculations of PDT
given in the textbook \cite{Landau8} in application to magnetic domains. The
geometry of the system is shown in Fig.\,\ref{FigSuplGeom}. The $x$-axis is
directed along the PDT texture, the $y$-axis is parallel to the domains and
the $z$-axis is directed across the film. Since the polarization
distribution does not change along the $y$-axis, the problem is reduced to
the 2D one in the $xz$-plane. In the external field the up- and
down-oriented polarization domains have different widths, $d_{+}$ and $d_{-}$%
, the field-induced polarization due to DW motion being expressed via the
field dependence of the asymmetry factor $\eta =\left( d_{+}-d_{-}\right)
/2d $ as $P_{\mathrm{dw}}=\eta (D)P_{\mathrm{s}}$, with $2d=d_{+}+d_{-}$
being the period of the structure. Following the relation (\ref{epsf1}) we
express the dielectric permittivity of the PDT as: 
\begin{equation}
\varepsilon _{\mathrm{f}}=\frac{\varepsilon _{\parallel }}{1-\left( \eta
(D)/D\right) P_{\mathrm{s}}}.  \label{ef}
\end{equation}%
The linear dependence $\eta $ on $D$ is found by minimization of the
electrostatic energy of the system, considering $\eta $ as variation
parameter.


We start with the derivation of the spacial distribution of the electric
fields inside the ferroelectric ($f$) and paraelectric ($p$) layers, $%
\mathbf{E}^{(p,f)}=(E_{x}^{(p,f)},E_{z}^{(p,f)})$, induced by the
depolarization surface charges arising at the interfaces between $f$- and $p$%
- layers at $z=\pm a_{\mathrm{f}}$. These charges, appearing due to
termination of the alternating polarization in PDT, are conventionally
presented as $\sigma _{\pm }(x)=\pm \vartheta (x)P_{\mathrm{s}}$ where the
sign-alternating function, $\vartheta $, is defined as $\vartheta (x)=\pm 1$
if $x\in d_{\pm }$. The corresponding electrostatic potentials in the $f$-
and $p$-layers $\nabla \varphi ^{(p,f)}=-\mathbf{E}^{(p,f)}$ satisfy the
Laplace equations: 
\begin{gather}
(\varepsilon _{\parallel }\partial _{z}^{2}+\varepsilon _{\perp }\partial
_{x}^{2})\varphi ^{(f)}=0,  \label{ElT} \\
\varepsilon _{\mathrm{p}}(\partial _{z}^{2}+\partial _{x}^{2})\varphi
^{(p)}=0.  \notag
\end{gather}%
The boundary conditions for the potential at the interfaces are fixed by the
charges $\sigma _{\pm }(x)$: 
\begin{gather}
\varepsilon _{0}\varepsilon _{\parallel }\partial _{z}\varphi
^{(f)}-\varepsilon _{0}\varepsilon _{\mathrm{p}}\partial _{z}\varphi
^{(p)}=\pm \sigma _{\pm }(x),  \label{bcT} \\
\varphi ^{(f)}=\varphi ^{(p)}.  \notag
\end{gather}%
System (\ref{ElT}) with boundary conditions (\ref{bcT})\textbf{\ } is solved
by the Fourier-series expansion method, taking into account that 
\begin{equation}
\sigma _{\pm }(x)=\pm \vartheta (x)P_{\mathrm{s}}=\pm \left( \eta
+\sum_{n=1}^{\infty }p_{n}\cos q_{n}x\right) P_{\mathrm{s}}  \label{PFour}
\end{equation}%
with $q_{n}=\pi n/d$ and 
\begin{equation}
p_{n}=\frac{4}{\pi n}\sin \frac{1+\eta }{2}\pi n  \label{qPn}
\end{equation}%
After straightforward calculations we obtain: 
\allowdisplaybreaks 
\begin{align}
\varphi ^{(f)}& =\frac{z}{a_{\mathrm{f}}}\varphi _{0}+P_{\mathrm{s}%
}\sum_{n=1}^{\infty }\psi _{n}\,\frac{\sinh \left( \sqrt{\varepsilon _{\perp
}/\varepsilon _{\parallel }}\,q_{n}z\right) }{\sinh \left( \sqrt{\varepsilon
_{\perp }/\varepsilon _{\parallel }}\,q_{n}a_{\mathrm{f}}\right) }\,\cos
q_{n}x,  \label{phis} \\
\varphi ^{(p)}& =\pm \varphi _{0}-\left( z-a_{\mathrm{f}}\right) \frac{D}{%
\varepsilon _{0}\varepsilon _{\mathrm{p}}}  \notag \\
& \pm P_{\mathrm{s}}\sum_{n=1}^{\infty }\,\psi _{n}\exp q_{n}(a_{\mathrm{f}%
}-\left\vert z\right\vert )\,\cos q_{n}x  \notag
\end{align}%
where $\varphi _{0}=\mp \left( \varepsilon _{0}\varepsilon _{\parallel
}\right) ^{-1}\left( D-\eta P_{\mathrm{s}}\right) a_{\mathrm{f}}$ is the
average value of the potential at the upper/lower surface of the
ferroelectric slab, and 
\begin{equation}
\psi _{n}=\frac{p_{n}}{q_{n}\,\varepsilon _{0}\sqrt{\varepsilon _{\perp
}\varepsilon _{\parallel }}\,\coth \left( \sqrt{\varepsilon _{\perp
}/\varepsilon _{\parallel}}\,q_{n}a_{\mathrm{f}}\right) +q_{n}\,\varepsilon
_{0}\varepsilon _{\mathrm{p}}}.  \label{psi}
\end{equation}
The electrostatic energy of the system can be calculated as the integral $%
\sum_{\pm }\int dx\int \left[ \varphi (x)\right] _{z=\pm a_{\mathrm{f}%
}}d\sigma _{\pm }(x)\,$ over the surfaces of the ferroelectric layer. The
integration over $\sigma _{\pm }$ reflects the self-consistent adiabatic
charging of the surfaces, driven by the self-polarization of the system from
the paraelectric state, until the state with the equilibrium spontaneous
polarization, $\pm P_{\mathrm{s}}$, is achieved inside domains.

Taking into account the dependicies given by (\ref{phis}) of $\varphi (x)$
and $\sigma _{\pm }(x)$ on $P_{\mathrm{s}}$ we calculate the energy per unit
length of PDT ( here the factor 2 is because of the two 2 sides of the
slab): 
\begin{widetext}
\begin{eqnarray}
F_{\mathrm{el}} &=&2\frac{1}{2d}\int_{0}^{2d}dx\int_{0}^{P_{\mathrm{s}}}%
\left[ \varphi (x,P_{\mathrm{s}}^{\prime })\right] _{z=a_{\mathrm{f}}}
\vartheta (x) dP_{\mathrm{s}}^{\prime }=2\int_{0}^{P_{\mathrm{s}}}dP_{%
\mathrm{s}}^{\prime }\left( \varphi _{0}(P_{\mathrm{s}}^{\prime })\eta +%
\frac{1}{2}P_{\mathrm{s}}^{\prime }\sum_{n=1}^{\infty }p_{n}\psi _{n}\right)
\label{Fel} \\
&=&\frac{2a_{\mathrm{f}}}{\varepsilon _{0}\varepsilon _{\parallel }}\left( 
\frac{1}{2}P_{\mathrm{s}}^{2}\eta ^{2}-DP_{\mathrm{s}}\eta \right) +P_{%
\mathrm{s}}^{2}\sum_{n=1}^{\infty }\frac{8d}{\pi ^{3}n^{3}}\frac{\sin
^{2}(1+\eta )\pi n/2}{\varepsilon _{0}\sqrt{\varepsilon _{\perp }\varepsilon
_{\parallel }}\coth \left( \pi n\sqrt{\varepsilon _{\perp }/\varepsilon
_{\parallel }}\,a_{\mathrm{f}}/d\right) +\varepsilon _{0}\varepsilon _{%
\mathrm{p}}}\,,  \notag
\end{eqnarray}%
\end{widetext}
In the experimentally relevant limit $\sqrt{\varepsilon _{\perp
}/\varepsilon _{\parallel }}\,a_{\mathrm{f}}/d\gg 1$, Eq.~(\ref{Fel}) is
simplified as: 
\begin{equation}
F_{\mathrm{el}}=\frac{2a_{\mathrm{f}}}{\varepsilon _{0}\varepsilon
_{\parallel }}\left( \frac{1}{2}P_{\mathrm{s}}^{2}\eta ^{2}-DP_{\mathrm{s}%
}\eta \right) +\frac{8}{\pi ^{3}\varepsilon _{0}\varsigma \sqrt{\varepsilon
_{\perp }\varepsilon _{\parallel }}}df(\eta )P_{\mathrm{s}}^{2},
\label{Fellshort}
\end{equation}%
where 
\begin{equation}
f(\eta )=\sum_{n=1}^{\infty }n^{-3}\sin ^{2}(1+\eta )\frac{\pi n}{2}\overset{%
\eta \rightarrow 0}{\simeq }\frac{7}{8}\zeta \left( 3\right) -\frac{\ln 2}{4}%
\left( \pi \eta \right) ^{2}.  \label{fdelta}
\end{equation}%
Minimization of (\ref{Fellshort}) with respect $\eta $ gives: 
\begin{equation}
\eta =\frac{D}{P_{\mathrm{s}}}\left( 1-\frac{4\ln 2}{\pi \varsigma }\left( 
\frac{\varepsilon _{\parallel }}{\varepsilon _{\perp }}\right) ^{1/2}\frac{d%
}{2a_{\mathrm{f}}}\right) ^{-1}.  \label{delta}
\end{equation}%
Finally, making use (\ref{ef}) we obtain the dielectric permittivity of the
PDT\,(\ref{Epsstat}).

The DW displacements in PDT shown in Fig.\,\ref{FigOscil}d are given by $\pm
x=\pm \eta d/2$. Equating the stiffness energy of the DW displacement, $2a_%
\mathrm{f}\cdot kx^{2}/2$, to (\ref{Fellshort}) (taken in harmonic
approximation at $D=0$ and multiplying by $d$, i.e. using the energy per a
single DW), we obtain the expression for the stiffness coefficient, $k$,\,(%
\ref{stiffness}).

Note that the expression\,(\ref{Epsstat}) can also be derived from the
formulas given in \cite{Bratkovsky2001} by expressing the ratio $P_{a}/U$ in
Eq.\,(31) of \cite{Bratkovsky2001} via the ratio $R/S$ defined in their Eq.\,(25). The Landau-Kittel formula (\ref{LK}) for the domain width, $d$, is
obtained by minimization of $F_{\mathrm{el}}+F_{\mathrm{dw}}$ energy with
respect $d$ at $\eta =0$. Here $F_{\mathrm{dw}}=(a_{f}/d)w_{\mathrm{dw}}$ is
the \ domain wall energy, calculated per unit length of PDT and $w_{\mathrm{%
dw}}$ is the surface energy of DW. The spatial distribution of the electric
and polarization fields in the $p$- and $f$- slabs are found from (\ref{phis}%
) as: $\mathbf{E}^{(p,f)}=(E_{x}^{(f)},E_{z}^{(f)})=-\nabla \varphi ^{(p,f)}$%
, $\mathbf{P}^{(p)}=\varepsilon _{0}\left( \varepsilon _{\mathrm{p}%
}-1\right) \mathbf{E}^{(p)}$, $\mathbf{P}^{(f)}=\left(
P_{x}^{(f)},P_{z}^{(f)}\right) $ where $P_{x}^{(f)}=\varepsilon _{0}\left(
\varepsilon _{\perp }-1\right) E_{x}^{(f)},P_{z}^{(f)}=\varepsilon
_{0}\left( \varepsilon _{\parallel }-1\right) E_{z}^{(f)}+\vartheta (x)P_{%
\mathrm{s}}$.
For the very thin films the field lines of $\mathbf{P}((r)$ form the closed loops, resembling the alternating vortex-antivortex patterns \cite{Yadav2016}

\bibliographystyle{apsrev4-1}
%
\end{document}